\documentclass[prl,twocolumn,superscriptaddress]{revtex4}
\usepackage{latexsym,amssymb,graphicx,caption}
\usepackage{epsfig}

\newcommand{\be}{\begin{equation}}
\newcommand{\ee}{\end{equation}}
\newcommand{\bea}{\begin{eqnarray}}
\newcommand{\eea}{\end{eqnarray}}

\def\figone#1#2#3{\begin{figure}
\centering \leavevmode \epsfxsize=0.8\columnwidth \epsfbox{#1}
\caption{#2 \label{#3}}
\end{figure} }

\begin{document}
\title{Landau theory of glassy dynamics}
\author{Satya N. Majumdar}
\affiliation{Laboratoire de Physique Theorique et Modeles
Statistiques, Universite Paris-Sud, Bat 100, 91405, Orsay-Cedex,
France} \affiliation{Laboratoire de Physique Theorique (UMR C5152
du CNRS), Universite Paul Sabatier, 31062 Toulouse Cedex, France}
\author{Dibyendu Das}\affiliation{Department of Physics, Indian Institute of Technology Bombay,
Powai, Mumbai 400076, India}
\author{Jan\'e Kondev}\author{Bulbul Chakraborty}
\affiliation{Martin Fisher School of Physics, Brandeis University,
Waltham, MA 02454 }

\begin{abstract}
An exact solution of a Landau model of an order-disorder
transition with {\em activated} critical dynamics is presented.
The model describes a funnel-shaped topography of the order
parameter space in which the number of energy lowering
trajectories rapidly diminishes as the ordered ground-state is
approached. This leads to an asymmetry in the effective transition
rates which results in a non-exponential relaxation of the
order-parameter fluctuations and a Vogel-Fulcher-Tammann
divergence of the relaxation times, typical of a glass transition.
We argue that the Landau model provides a general framework for
studying glassy dynamics in a variety of systems.
\end{abstract}


\maketitle

\paragraph{Introduction}Glassy dynamics occur in a large variety
of systems, such as supercooled liquids, foams and granular
matter. They are characterized by an extremely rapid increase of
relaxation times and by a non-exponential decay of time-dependent
correlation functions\cite{GlassRev1}. The rapid increase in time
scales is typically fit by an exponential, Vogel-Fulcher-Tammann
(VFT) dependence on a control parameter such as temperature or
density. In this letter, we propose a Landau model of
order-parameter dynamics in the vicinity of a critical point,
which exhibits these features.

The canonical theoretical framework for dynamics near a critical
point is the time-dependent Ginzburg-Landau
equation\cite{Goldenfeld}. It successfully describes the
phenomenon of critical slowing down, whereby the relaxation time
for order-parameter fluctuations scales as a power of the
correlation length, which in turn diverges as the critical point
is approached.  A natural question to ask is whether this
framework can be adapted to describe glassy dynamics. Here we show
analytically that a master equation, based on a Landau free
energy, where the energy and entropy play asymmetric roles in
determining the transition rates, naturally leads to a VFT
divergence and a broad distribution of relaxation time scales near
a critical point.

The microscopic basis for the  asymmetry in the transition rates
can be found in the topology of the space of trajectories. Namely,
we consider the situation where the number of energy lowering
trajectories diminishes as the ordered ground state is approached.
A similar funnel-like topology of the trajectory space for a
simple kinetic model of protein folding has been identified
recently as a possible explanation of fast folding\cite{DillPNAS}.
Here we show that such a topology of the trajectory space, when a
system is driven to a critical point, can result in glassy
critical dynamics. Furthermore, we hypothesize that diverse
systems such as foams, granular matter and supercooled liquids owe
their glassy features to such a mechanism, and propose
computational tests of this idea.

\paragraph{Model}We start with a mean-field model of a continuous phase
transition for a scalar order parameter $\rho$ which takes values on the
integers. We describe dynamics of $\rho$ as a random
walk on the order parameter space which is a one-dimensional lattice.
The Landau free energy in the disordered phase ($\mu<\mu^*$),  is
assumed to be quadratic
\begin{equation}
\beta F(\rho) = \beta F^0+ {\mu^*-\mu\over 2} \rho^2 \ ;
\label{freeenergy}
\end{equation}
$\mu$ is the dimensionless control parameter, and
$\mu^*$ is its critical value.
The energy and the entropy are also assumed to be quadratic
functions of the order parameter: $\beta E(\rho) = \beta E^0-
\mu\rho^2/2$, and $S(\rho) =  S^0 -
\mu^*{\rho}^2/2$\cite{footnote1}. The free energy $\beta F = \beta
E - S$ leads to a mean-field phase transition as ${\mu}^* -\mu
\rightarrow 0$. Below we shall be interested in the order
parameter dynamics as the phase transition point is approached from the
disordered phase.

In the Landau approach to critical dynamics, $P(\rho, t)$, the
probability for the order parameter to take the value $\rho$ at time $t$,
is the solution to the master equation
\bea
\label{diseqn1}
{\partial P(\rho) \over \partial t} & = & - (W_{\rho\to\rho-1}+W_{\rho\to\rho+1})P(\rho) \nonumber \\
& & + W_{\rho-1,\rho} P(\rho-1) + W_{\rho+1\to\rho}P(\rho+1) \ .
\eea The transition rates, $W_{\rho\to\rho'}$, depend on the free
energy, the canonical choice being, $W_{\rho\to\rho'}=\exp[-(\beta
F(\rho')-\beta F(\rho))/2)]$. These rates satisfy detailed balance
and yield the equilibrium Gibbs distribution, $P_{\rm eq}(\rho)
\propto \exp[-\beta F(\rho)]$, as the steady state solution to
Eq.(\ref{diseqn1}). Relaxation processes described by these rates
implicitly assume that neither energy nor entropy barriers impede
the fluctuations of the order parameter around its equilibrium
value, $\rho_{\rm eq}=0$.
\begin{figure}
\epsfbox{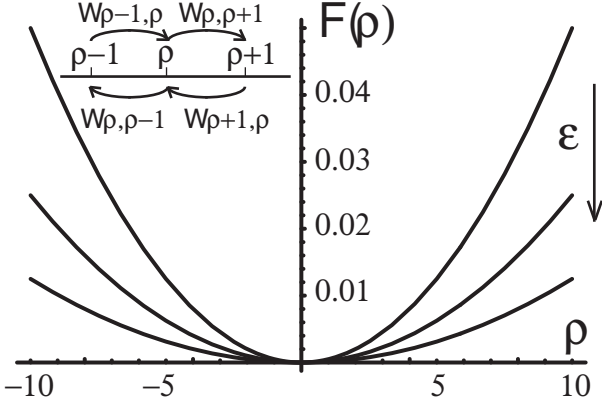}\caption{The Landau free energy, $\beta
F(\rho)$ is a parabola whose width increases as the transition is
approached ($\epsilon\to 0$). The order parameter $\rho$ performs
a random walk with rates dictated by the energy and the entropy,
which are functions of $\rho$.}\label{fig:freeEnergy}
\end{figure}

The canonical scenario described above can be altered dramatically
if the microscopic dynamics are constrained in some way. An
example is provided by a system of hard spheres at densities
approaching random close packing. At these large densities only
correlated motions of many spheres are available to the system in
order to relax a density perturbation\cite{Torquato}. We consider
the case where such constraints in the microscopic dynamics can be
encoded as an asymmetry in the way energy and entropy enter the
transition rates $W_{\rho\to\rho'}$. In particular, we assume that
energy lowering transitions can proceed only from a small subset
of microstates at given $\rho$. If the system finds itself in one
of these microstates the transition occurs at a fixed rate
independent of the value of $\rho$. Thus the rate of energy
lowering transitions is a function of the {\it entropy} change
only. Taking the requirement of detailed balance also into account
leads to \bea \label{rate:model}
W_{\rho\to\rho'} &  = &  e^{-[S(\rho)-S(\rho')]} ~; E(\rho')<E(\rho) \nonumber \\
W_{\rho\to\rho'} & = & e^{-[\beta E(\rho')-\beta E(\rho)]}~; E(\rho')>E(\rho) \ .
\eea


Using the quadratic expressions for the energy and entropy in these
transition rates leads to
 \bea \label{diseqn2}
{\partial P(\rho) \over \partial t} & = & - \left( e^{-\mu\rho} + e^{\mu^*(\rho+1)} \right) P(\rho) \nonumber \\
 & + & e^{-\mu^*\rho} P(\rho-1) + e^{-\mu(\rho+1)}
P(\rho+1) \eea for the master equation for $P(\rho,t)$ for $\rho
>0$. Since the rates are symmetric, the equation for $\rho <0$ has a similar form.
Close to the critical point ($\mu^* - \mu \ll 1$) typical
values of $\rho$ are much greater than one, and we can take the
continuum-$\rho$ limit of this equation to obtain\cite{foot_x0}:
\be {\partial P \over \partial t} = {\partial \over \partial
\rho}[\epsilon \rho e^{-{\mu }^* |\rho|} P + 2 D e^{-{\mu }^*
|\rho|} {\partial P \over
\partial \rho}] \label{fp1}
\ee
with $\epsilon=\mu^*-\mu$ the distance to the transition
point, and $D=1/2$.
We note that a direct consequence of the
asymmetric transition rates is the exponentially decaying factor
multiplying both the drift and the diffusion terms
in Eq.(\ref{fp1}). The time-invariant probability distribution
obtained from this equation is the normalized, equilibrium
distribution: $P_{eq}(\rho)= \sqrt{ {\epsilon}\over {4\pi}}
e^{-{{\epsilon}\over {4D}} {\rho}^2}$.


\paragraph{Scaling solution at the critical point} For
$\epsilon=0$,  the Laplace transform ${\tilde
P}(\rho,s)=\int_0^{\infty}P(\rho,t)e^{-st}dt$ is given exactly by,
\begin{equation}
{\tilde P}(\rho,s)= {1\over {\sqrt{8sD}}}e^{{\mu}^*|\rho|/2} {
{K_1\left(\sqrt{ {2s}\over {D{\mu}^{*2}}}e^{{\mu}^*
|\rho|/2}\right)}\over { K_0\left(\sqrt{ {2s}\over
{D{\mu}^{*2}}}\right)} }. \label{lt1}
\end{equation}

The long-time limit $t\to \infty$ corresponds to the $s\to 0$
limit in Laplace space and Eq.(\ref{lt1})
suggests the scaling limit:  $s\to 0$ and
$|\rho|\to \infty$ while keeping $\sqrt{s} e^{{\mu}^* |\rho|/2}$ fixed.
Using the  scaling variable $z = \frac{e^{{\mu}^*
|\rho|/2}}{\sqrt{t}}$, the scaling form of the solution to Eq.(\ref{fp1}) can
be written as
\begin{equation}
P(\rho,t) =  {{\mu}^* \over {2\log Dt}} F(z) \ .
\label{psol1}
\end{equation}
The scaling function $F(z)= {{\mu}^* \over 2}
e^{-z^2/{2{\mu}^{*2}}}$ is determined by substituting
the scaling form, Eq.(\ref{psol1}), in Eq.(\ref{fp1}) with $\epsilon=0$, and
matching the leading order terms for large $t$.
This solution immediately gives us the mean square fluctuation of
$\rho$  for large $t$,
\begin{equation}
\langle {\rho}^2(t)\rangle ={ {\int_0^{\infty} {\rho}^2
\exp\left[- {{e^{{\mu}^*{\rho}}}\over {2{\mu}^{*2} Dt}}\right]
d\rho} \over {\int_0^{\infty} \exp\left[- {{e^{{\mu}^*\rho}}\over
{2{\mu}^{*2} Dt}}\right] d\rho}} \sim
\log^2 (2{\mu^*}^2 Dt). \label{ms1}
\end{equation}
The fluctuations grow logarithmically with time, in contrast to
the standard $\sqrt{t}$ scaling expected at a normal critical
point.

\paragraph{Scaling solution for $\epsilon \ne 0$}Away from the
critical point, the Fokker-Planck equation, Eq.(\ref{fp1}), has a
normalizable stationary solution. The equation obtained
by substituting $P(\rho,t)= e^{-\epsilon {\rho}^2/{4D}}\psi(\rho
,t)$ in Eq.(\ref{fp1}) allows a scaling solution for $\psi(\rho ,
t)$ in the limit $t\to \infty$, $\rho \to \infty$ but keeping the
combination $e^{{\mu}^* \rho}/t$ fixed. Repeating the same
steps as described above for the $\epsilon=0$ case we compute
\begin{equation}
P(\rho,t) = { {\exp{(-\epsilon {\rho}^2/{4D}- {e^{{\mu}^*|\rho|}
\over {2{\mu}^{*2} Dt }})} }\over { 2\int_0^{\infty} d\rho\,
\exp{(-\epsilon {\rho}^2/{4D}- {e^{{\mu}^* \rho} \over
{2{\mu}^{*2} Dt }})} } }. \label{sc4}
\end{equation}
\figone{prob_disbn_log.eps}{The probability distribution, $P(\rho
,t)$  for three different times,  with $\epsilon = 0.1$, $D=1.$,
and $\mu^* = 1$. For these parameters, ${\rho}_c = 0.69,~2.3,~4.6$
for $t=2,~10,~100$, respectively, and $t_c = 88$.  Note that at
$t=2$ the distribution is essentially super-exponential and for
$t=100$ it is a Gaussian.}{fig:probdisbn} In the limit $t\to
\infty$, this solution recovers the equilibrium distribution.
{}From Eq.(\ref{sc4}), it is clear that at a given large $t$,
there are two length scales in $\rho$-space: the equilibrium
length scale,$\sqrt{\langle {\rho}^2\rangle_{eq}}=
\sqrt{2D/\epsilon}$, and a time-dependent length scale, $\rho_c =
(1/{\mu}^*) \log({\mu}^{*2} D t)$. For $\rho \ll \rho_c$,
$P(\rho,t)$ reduces to the equilibrium Gaussian distribution while
for $\rho \gg \rho_c$ the distribution of $\rho$ is
super-exponential. At long times $\rho_c$ becomes much larger than
the equilibrium $\rho$ and the system ``knows'' that it is in
equilibrium. This defines a crossover time $t_c \sim
e^{{\mu}^*\sqrt{2D/\epsilon}}$ obtained by equating $\rho_c$ with
$\sqrt{\langle {\rho}^2\rangle_{eq}}$.
Below
we also compute the relaxation time from the the mean-square
fluctuation of the order parameter, $\left<\rho^2(t)\right>$ and
show that it diverges as $\exp[1/\epsilon]$, which is precisely
the VFT law. An interesting conclusion that can be drawn about the
Landau model is that its critical dynamics
cannot be characterized by a single diverging
time scale.



We have seen earlier that in the $\epsilon=0$ case, $\langle
{\rho}^2(t)\rangle \sim \log^2(t)$ for large $t$. For $\epsilon\ne
0$ this quantity will relax to its equilibrium value $\langle
{\rho}^2\rangle_{eq}= 2D/\epsilon$. From Eq.(\ref{sc4}), we find
in the scaling limit,
\begin{equation}
\langle {\rho}^2(t)\rangle = { \int_0^{\infty}
{\rho}^2{\exp{(-\epsilon {\rho}^2/{4D}- {e^{{\mu}^* {\rho}} \over
{2{\mu}^{*2} Dt}})}d{\rho} }\over { \int_0^{\infty} \exp{(-\epsilon
{\rho}^2/{4D}- {e^{{\mu}^* {\rho}} \over {2{\mu}^{*2} Dt}})}
d{\rho}} }. \label{ms2}
\end{equation}
Thus one can write,
\begin{equation}
\langle {\rho}^2(t)\rangle = -{{\partial \log[Z(A)]}\over
{\partial A}}, \label{ms3}
\end{equation}
where $A= \epsilon/{4D}$ and the generating function $Z(A)$ is
given by,
\begin{equation}
Z(A) = \int_0^{\infty}d\rho\, \exp{\left( -A {\rho}^2- {e^{{\mu}^* \rho
}\over {2{\mu}^{*2} Dt}}\right)}. \label{pf1}
\end{equation}
From Eq.(\ref{ms3}) we find
that for large $t$ and small $A$,
\begin{equation}
\langle {\rho}^2(t)\rangle \approx \langle {\rho}^2\rangle_{eq} -
{{4D}\over {\epsilon^2} } e^{D{\mu}^{*2}/\epsilon} {1\over {t}} \ .
\label{ms4}
\end{equation}
Thus, at late times, the mean square fluctuation relaxes  to its
equilibrium value in a power law ($\sim 1/t$) fashion. A standard
way to estimate the relaxation time $\tau$ is to define it as the
time needed for the difference $\langle
{\rho}^2\rangle_{eq}-\langle {\rho}^2(t)\rangle$ to decay to a
given value of order one.  {}From Eq.(\ref{ms4}) we immediately
obtain
\begin{equation}
\tau \propto {{D}\over {\epsilon^2} }e^{D{\mu}^{*2}/\epsilon},
\label{vf1}
\end{equation}
which diverges in the VFT manner as $\epsilon\to 0$.

Activated dynamics at a critical point were previously shown to
occur in systems with quenched disorder\cite{Fisher} and it had
been suggested that a similar situation might exist in
non-disordered systems with frustration\cite{Sethna}. The
emergence of a glass state in the latter case could then be
associated with a critical point at which relaxation times would
acquire astronomical values long before any discernible spatial
correlations indicative of an ordered state could be established.
The dynamical Landau model presented here provides an explicit
realization of this scenario.


The scaling solution to our model demonstrates that a seemingly
innocuous change in the transition rates leads to an exponential
divergence of timescales at a critical point, while all its static
properties are described by mean-field Landau theory.  The
dynamics of the Landau model are characterized by a multitude of
timescales and by non-exponential relaxations even away from the
critical point. Namely, the mean-square fluctuation of $\rho$ can
be written in terms of the density of states $\Omega (E)$ as:
$\langle{\rho}^2(t)\rangle = \int dE \Omega(E) e^{-Et}$.  The
power law decay for $\epsilon \ge 0$, Eq.(\ref{ms4}), implies that
$\Omega(E)$ is proportional to  $\delta (E) - {\rm const}$ with
the $\delta$-function guaranteeing the equilibrium value.  At the
critical point, the logarithmic decay of the fluctuations implies
$\Omega(E) \propto \ln^2(E)/E$. The glass transition in our model
is, therefore, not characterized by  a vanishing gap in the energy
spectrum (of Eq.~\ref{fp1} ) but  instead by
an accumulation of states near $E=0$.  This then translates to a
distribution of timescales with increasing weight in the tails of
the distribution corresponding to long times.

In the Landau model with a quadratic entropy function the entropy
goes to zero at ${\rho}_{max}^2 = 2 S_0/{\mu}^*$ and our analysis
is, therefore, strictly valid for $\langle {\rho}^2\rangle_{eq}
\le {\rho}_{max}^2$ or $\epsilon \ge {\mu}^*/2 S_0$.
For smaller values of $\epsilon$ the relaxation time-scale
divergence is cut off by the system size.

\paragraph{Microscopic Models}
The transition rates given by Eq.(\ref{rate:model}), which formed the
basis of our dynamical equation and led to the VFT divergence of
timescales, are explicitly realized in the interacting three-color
model\cite{Dibyendu}.  The configurations of the three color model
on the honeycomb lattice can be mapped on to loop packings and the
loops  can be identified with equal height contours of a discrete
height field which lives on the dual triangular
lattice\cite{Kondev_Nienhaus_Grier}. Coloring configurations can be
organized into topologically distinct sectors characterized by
the number of non-zero winding number loops, which correspond to different global
tilts in the height representation. The introduction of a
long-range interaction between one of the colors leads to a phase
transition from a flat to a completely tilted state\cite{Dibyendu}
with the tilt playing the role of the order parameter $\rho$.

Microscopic  dynamics for the three color model can be defined in
terms of loop updates\cite{Dibyendu} with transitions between
different loop configurations satisfying detailed balance.
Simulations were performed to extract the transition rates between
different tilt states, $W_{\rho \to \rho'}$ and it was shown,
numerically, that they take the form described by
Eq.(\ref{rate:model}).  The asymmetry in the rates owes its origin
to the loop dynamics.  A $\rho$-reducing (energy-increasing)
transition involves deleting an existing non-zero winding number
loop.  The rate of this transition is determined {\it only} by the
change in {\it energy} since all loop configurations at a given
$\rho$ can lead to a configuration with $\rho' <\rho$. On the
other hand, a transition that increases $\rho$ and consequently
decreases the energy,  involves introducing a new non-zero winding
number loop.
Since only a small  subset
of loop configurations at a given $\rho$ can accommodate
a new non-zero winding number loop, this
transition rate  depends {\it only} on the change of {\it
entropy}. We believe similar mechanisms are operative in other
system where transitions between macro-states involve large
scale rearrangements which, due to the presence of constraints,
can occur only through coordination of many microscopically
rearranging regions.

In this work, we identify a concrete physical mechanism by which
dynamical barriers are established in the free energy landscape.
The dynamics in the order parameter space in our model is thus
reminiscent of diffusion on hierarchical
lattices\cite{Teitel,Palmer_Anderson} and trap
models\cite{bouchaud}. Recently, attempts have been made to
identify inherent structures in supercooled liquids as providing a
microscopic basis for traps\cite{Reichman,Heuer}. If we identify
inherent structures with the order parameter $\rho$ in our model
then the predicted distribution of trapping times, $\tau_{\rho} =
1/(W_{\rho\to\rho-1}+W_{\rho\to\rho+1})$,  is
log-normal with a width that increases as $1/{\epsilon}$ . This is
consistent with the behavior observed in supercooled liquid
simulations\cite{Reichman}.

Studies of kinetically constrained models\cite{Sollich_Ritort}
have identified dynamical
heterogeneities\cite{Garrahan3,Garrahan1} resulting from the
kinetic constraints and an entropy crisis in trajectory space
which leads to rapidly increasing
timescales\cite{Garrahan1,biroli}. In our model, the asymmetry of
the rates results in an entropy crisis and, combined with the
existence of an equilibrium critical point, leads to an
exponential divergence of relaxation time-scales for a finite
value of the control parameter.


The coarse-grained dynamical model presented here suggests a new
way of analyzing atomistic simulations of glassy materials. From
molecular dynamics simulations one can measure the transition
rates between macro-states characterized by different values of a
slow variable which plays the role of an order parameter, for
example, the metabasin energies in supercooled
liquids\cite{Reichman,Heuer}. This approach avoids making specific
assumptions about the origin of the transition rates. Dynamics in
order parameter space can be constructed explicitly and analyzed
for the types of asymmetries and related dynamical barriers
discussed here within a simple Landau model.  Work along these
lines is currently in progress.





We would like to thank  Albion Lawrence and Leticia Cugliandolo
for useful discussions. We acknowledge support by NSF grants
DMR-0207106 (BC) and DMR-9984471 (JK) and by the Research
Corporation through the Cottrell Scholar program (JK).

\bibliography{glass}

\begin{thebibliography}{19}
\expandafter\ifx\csname natexlab\endcsname\relax\def\natexlab#1{#1}\fi
\expandafter\ifx\csname bibnamefont\endcsname\relax
  \def\bibnamefont#1{#1}\fi
\expandafter\ifx\csname bibfnamefont\endcsname\relax
  \def\bibfnamefont#1{#1}\fi
\expandafter\ifx\csname citenamefont\endcsname\relax
  \def\citenamefont#1{#1}\fi
\expandafter\ifx\csname url\endcsname\relax
  \def\url#1{\texttt{#1}}\fi
\expandafter\ifx\csname urlprefix\endcsname\relax\def\urlprefix{URL }\fi
\providecommand{\bibinfo}[2]{#2}
\providecommand{\eprint}[2][]{\url{#2}}

\bibitem[{\citenamefont{Ediger et~al.}(1996)\citenamefont{Ediger, Angell, and
  Nagel}}]{GlassRev1}
\bibinfo{author}{\bibfnamefont{M.~D.} \bibnamefont{Ediger}},
  \bibinfo{author}{\bibfnamefont{C.~A.} \bibnamefont{Angell}},
  \bibnamefont{and} \bibinfo{author}{\bibfnamefont{S.~R.} \bibnamefont{Nagel}},
  \bibinfo{journal}{J. Phys. Chem.} \textbf{\bibinfo{volume}{100}},
  \bibinfo{pages}{13200} (\bibinfo{year}{1996}).

\bibitem[{\citenamefont{Goldenfeld}(1992)}]{Goldenfeld}
\bibinfo{author}{\bibfnamefont{N.}~\bibnamefont{Goldenfeld}},
  \emph{\bibinfo{title}{Lectures on Phase Transitions and the Renormalization
  Group}} (\bibinfo{publisher}{Addison-Wesley}, \bibinfo{address}{New York,
  NY}, \bibinfo{year}{1992}).

\bibitem[{\citenamefont{Schonbrun and Dill}(2003)}]{DillPNAS}
\bibinfo{author}{\bibfnamefont{J.}~\bibnamefont{Schonbrun}} \bibnamefont{and}
  \bibinfo{author}{\bibfnamefont{K.~A.} \bibnamefont{Dill}},
  \bibinfo{journal}{Proc. Nat. Acad. Sci.} \textbf{\bibinfo{volume}{100}},
  \bibinfo{pages}{12678} (\bibinfo{year}{2003}).

\bibitem[{foo({\natexlab{a}})}]{footnote1}
\bibinfo{note}{We will show below that by taking limits appropriately, the
  entropy does not become negative}.

\bibitem[{\citenamefont{Kansal et~al.}(2002)\citenamefont{Kansal, Torquato, and
  Stillinger}}]{Torquato}
\bibinfo{author}{\bibfnamefont{A.~R.} \bibnamefont{Kansal}},
  \bibinfo{author}{\bibfnamefont{S.}~\bibnamefont{Torquato}}, \bibnamefont{and}
  \bibinfo{author}{\bibfnamefont{F.~H.} \bibnamefont{Stillinger}},
  \bibinfo{journal}{Phys. Rev. E} \textbf{\bibinfo{volume}{66}},
  \bibinfo{pages}{041109} (\bibinfo{year}{2002}).

\bibitem[{foo({\natexlab{b}})}]{foot_x0}
\bibinfo{note}{The $\rho=0$ point is included in the continuum equation; Satya
  N. Majumdar, unpublished}.

\bibitem[{\citenamefont{Fisher}(1986)}]{Fisher}
\bibinfo{author}{\bibfnamefont{D.~S.} \bibnamefont{Fisher}},
  \bibinfo{journal}{Phys. Rev. Lett.} \textbf{\bibinfo{volume}{56}},
  \bibinfo{pages}{416} (\bibinfo{year}{1986}).

\bibitem[{\citenamefont{Sethna et~al.}(1993)\citenamefont{Sethna, Shore, and
  Huang}}]{Sethna}
\bibinfo{author}{\bibfnamefont{J.~P.} \bibnamefont{Sethna}},
  \bibinfo{author}{\bibfnamefont{J.~D.} \bibnamefont{Shore}}, \bibnamefont{and}
  \bibinfo{author}{\bibfnamefont{M.}~\bibnamefont{Huang}},
  \bibinfo{journal}{Phys. Rev. B} \textbf{\bibinfo{volume}{47}},
  \bibinfo{pages}{14661} (\bibinfo{year}{1993}).

\bibitem[{\citenamefont{Das et~al.}(2003)\citenamefont{Das, Kondev, and
  Chakraborty}}]{Dibyendu}
\bibinfo{author}{\bibfnamefont{D.}~\bibnamefont{Das}},
  \bibinfo{author}{\bibfnamefont{J.}~\bibnamefont{Kondev}}, \bibnamefont{and}
  \bibinfo{author}{\bibfnamefont{B.}~\bibnamefont{Chakraborty}},
  \bibinfo{journal}{Europhys. Lett.} \textbf{\bibinfo{volume}{61}},
  \bibinfo{pages}{506} (\bibinfo{year}{2003}).

\bibitem[{\citenamefont{Kondev et~al.}(1996)\citenamefont{Kondev, deGier, and
  Nienhuis}}]{Kondev_Nienhaus_Grier}
\bibinfo{author}{\bibfnamefont{J.}~\bibnamefont{Kondev}},
  \bibinfo{author}{\bibfnamefont{J.}~\bibnamefont{deGier}}, \bibnamefont{and}
  \bibinfo{author}{\bibfnamefont{B.}~\bibnamefont{Nienhuis}},
  \bibinfo{journal}{J. Phys. A} \textbf{\bibinfo{volume}{29}},
  \bibinfo{pages}{6489} (\bibinfo{year}{1996}).

\bibitem[{\citenamefont{Teitel}(1988)}]{Teitel}
\bibinfo{author}{\bibfnamefont{S.}~\bibnamefont{Teitel}},
  \bibinfo{journal}{Phys. Rev. Lett.} \textbf{\bibinfo{volume}{60}},
  \bibinfo{pages}{1154} (\bibinfo{year}{1988}).

\bibitem[{\citenamefont{Palmer~{\it et al}}(1984)}]{Palmer_Anderson}
\bibinfo{author}{\bibfnamefont{R.~G.} \bibnamefont{Palmer~{\it et al}}},
  \bibinfo{journal}{Phys. Rev. Lett.} \textbf{\bibinfo{volume}{53}},
  \bibinfo{pages}{958} (\bibinfo{year}{1984}).

\bibitem[{\citenamefont{Monthus and Bouchaud}(1996)}]{bouchaud}
\bibinfo{author}{\bibfnamefont{C.}~\bibnamefont{Monthus}} \bibnamefont{and}
  \bibinfo{author}{\bibfnamefont{J.-P.} \bibnamefont{Bouchaud}},
  \bibinfo{journal}{J. Phys. A} \textbf{\bibinfo{volume}{29}},
  \bibinfo{pages}{3847} (\bibinfo{year}{1996}).

\bibitem[{\citenamefont{Denny et~al.}(2003)\citenamefont{Denny, Reichman, and
  Bouchaud}}]{Reichman}
\bibinfo{author}{\bibfnamefont{R.~A.} \bibnamefont{Denny}},
  \bibinfo{author}{\bibfnamefont{D.~R.} \bibnamefont{Reichman}},
  \bibnamefont{and} \bibinfo{author}{\bibfnamefont{J.~P.}
  \bibnamefont{Bouchaud}}, \bibinfo{journal}{Phys. Rev. Lett.}
  \textbf{\bibinfo{volume}{90}}, \bibinfo{pages}{025503}
  (\bibinfo{year}{2003}).

\bibitem[{\citenamefont{Doliwa and Heuer}(2003)}]{Heuer}
\bibinfo{author}{\bibfnamefont{B.}~\bibnamefont{Doliwa}} \bibnamefont{and}
  \bibinfo{author}{\bibfnamefont{A.}~\bibnamefont{Heuer}},
  \bibinfo{journal}{Phys. Rev. E} \textbf{\bibinfo{volume}{67}},
  \bibinfo{pages}{030501} (\bibinfo{year}{2003}).

\bibitem[{\citenamefont{Ritort and Sollich}(2003)}]{Sollich_Ritort}
\bibinfo{author}{\bibfnamefont{F.}~\bibnamefont{Ritort}} \bibnamefont{and}
  \bibinfo{author}{\bibfnamefont{P.}~\bibnamefont{Sollich}},
  \bibinfo{journal}{Adv. in Phys.} \textbf{\bibinfo{volume}{52}},
  \bibinfo{pages}{219} (\bibinfo{year}{2003}).

\bibitem[{\citenamefont{Garrahan and Chandler}(2003)}]{Garrahan3}
\bibinfo{author}{\bibfnamefont{J.~P.} \bibnamefont{Garrahan}} \bibnamefont{and}
  \bibinfo{author}{\bibfnamefont{D.}~\bibnamefont{Chandler}},
  \bibinfo{journal}{Proc. Nat. Acad. of Sci.} \textbf{\bibinfo{volume}{100}},
  \bibinfo{pages}{9710} (\bibinfo{year}{2003}).

\bibitem[{\citenamefont{Garrahan and Chandler}(2002)}]{Garrahan1}
\bibinfo{author}{\bibfnamefont{J.~P.} \bibnamefont{Garrahan}} \bibnamefont{and}
  \bibinfo{author}{\bibfnamefont{D.}~\bibnamefont{Chandler}},
  \bibinfo{journal}{Phys. Rev. Lett.} \textbf{\bibinfo{volume}{89}},
  \bibinfo{pages}{035704} (\bibinfo{year}{2002}).

\bibitem[{\citenamefont{Toninelli et~al.}(2003)\citenamefont{Toninelli, Biroli,
  and Fisher}}]{biroli}
\bibinfo{author}{\bibfnamefont{C.}~\bibnamefont{Toninelli}},
  \bibinfo{author}{\bibfnamefont{G.}~\bibnamefont{Biroli}}, \bibnamefont{and}
  \bibinfo{author}{\bibfnamefont{D.~S.} \bibnamefont{Fisher}},
  \bibinfo{journal}{cond-mat/0306746}  (\bibinfo{year}{2003}).

\end{thebibliography}
\end{document}